\documentclass[aps,prl,twocolumn,english,groupedaddress]{revtex4}
\usepackage{graphicx}
\usepackage{ amsmath, amssymb}

\usepackage{babel}
\makeatother
\begin{document}

\title{Ferromagnetic Relaxation by Magnon Induced Currents}

\author{A. Misra}
\email{misra013@umn.edu}
\author{R. H. Victora}
\email{victora@ece.umn.edu}

\affiliation{Department of Electrical and Computer Engineering, University of
Minnesota, Minneapolis, Minnesota 55455}

\begin{abstract}
A theory for calculating spin wave relaxation times based on the magnon-electron
interaction is developed. The theory incorporates a thin film geometry and 
is valid for a large range of magnon frequencies and wave vectors. For high
conductivity metals such as permalloy, the wave vector dependent damping constant
approaches values as high as $0.2$, 
showing the large magnitude of the effect, and can dominate experimentally observed relaxation.
\end{abstract}

\maketitle

One of the fundamental problems of magnetism is to determine a mechanism 
for dissipation of energy in a system subject to a change in the direction
of the  external magnetic field. 
Historically, these ferromagnetic relaxation processes have been explored by 
ferromagnetic resonance (FMR) in which the absorption of a small RF frequency
field applied perpendicular to a large DC field is measured. More recently,
direct measurement of large angle switching has been made by multiple groups
\cite{HSF1997PRL,Choi2001PRL,AB2000Sci}.
Time resolved Kerr microscopy has made it possible to track isolated magnetic
relaxation processes with picosecond temporal resolution \cite{HSF1997PRL} 
and also to
image the individual components of the precessing magnetization \cite{AB2000Sci}.
It has been previously shown that for many examples of large angle 
switching, particularly those involving materials with large magnetizations, the
dominant relaxation process is very different than that applicable to FMR.
In particular, it was found that the coherent mode scatters with thermal
magnons to form two $k\ne0$ magnons \cite{DV2003PRL}. This 4-magnon process rapidly escalates
as the $k\ne0$ magnon levels are populated, thus promoting additional scattering.
This previous work thus accounted for the rapid movement of the magnetization into
the new direction, but the dissipation of the magnetic energy stored in the
$k\ne0$ modes still must be addressed.
 
In a conducting ferromagnet the interaction between the conduction electrons
and the magnons become very important.  
The magnetic field generated by the spin wave is
time dependent and therefore, by Faraday's law, it creates an electric field 
in the system. These
electric fields, unlike conventional eddy currents, are wave like in nature.
In a metallic system, the fields drive the conduction electrons. These 
magnon induced currents
help dissipate the energy of the system by Joule heating.
Abrahams \cite{A1955PR}
addressed this question half a century ago by taking into account the 
interaction between spin waves and conduction electrons. 
However, his bulk estimates predicted relaxation times one order of
magnitude less than required to explain FMR line widths.  Subsequently 
several attempts have been made to give a consistent theory of ferromagnetic
relaxation from the point of view of an FMR experiment \cite{KP1975PRB}.
Later, Almeida and Mills \cite{AM1996PRB} explored the same interaction 
and derived the Green's function for the limited case of small angle precession
in the absence of quantum mechanical exchange, i.e. in the long wavelength
limit.

In the present work, we avoid much of the limitations found in previous work.
Solution of the general problem is difficult because the magnon-induced currents 
generate new fields which further affect magnons and create new currents. 
Here we show that expansion in the small parameter $4\pi\omega\sigma/
c^2k^2$ allows an explicit solution to the general problem. 
It allows prediction of decay rates for magnons of arbitrary frequency and 
large amplitudes that are limited only by the Holstein Primakoff transformation
 \cite{HP1940PR}.
In particular, it can be applied to the problem of magnons generated by 
four-magnon 
scattering after a large angle rotation such is common in modern switching
experiments and technological applications such as magnetic recording.
We also discuss our results in context of the spin wave resonance experiments
capable of measuring the linewidth of the higher order $k$ modes. Historically,
these experiments were used to measure the exchange constant of the material. 

\begin{figure}
\includegraphics[width=2.9in]{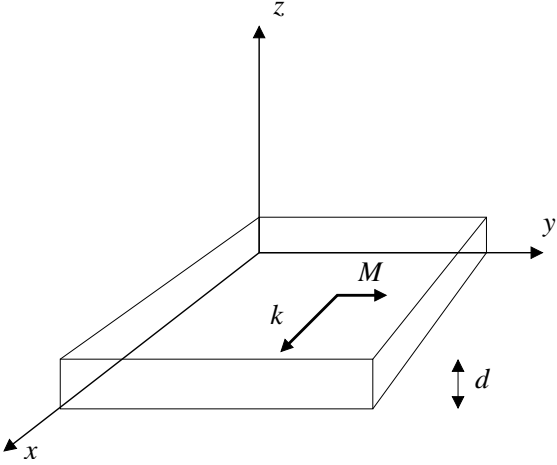}
\caption{\label{scheme}Schematic diagram showing the geometry used in the 
paper. An infinite ferromagnetic metallic slab with thickness $d$ along 
the $z$-direction is magnetized in-plane by applying an external field 
$\mathbf H$. We consider spin waves with wave vector $k$ propagating parallel
to the surfaces. }
\end{figure}

We consider an infinite film of thickness $d$ made of ferromagnetic
metal. The top and the bottom surfaces of the film are at $z=d$ and
$z=0$ respectively (see Fig. \ref{scheme}). A spin wave of wave vector 
$\mathbf{k}=k\widehat{x}$ and frequency $\omega$ 
is excited in the system. 

We write the electric and magnetic fields in the system as a series expansion
\begin{eqnarray}
\mathbf{E} & = & \sum_{n=0}^{\infty} \left( \frac{4\pi\sigma\omega}{c^2k^2} 
\right)^n \mathbf{E}^{(n)} \nonumber \\
\mathbf{H} & = & \sum_{n=0}^{\infty} \left( \frac{4\pi\sigma\omega}{c^2k^2} 
\right)^n \mathbf{H}^{(n)} \label{eq:series}
\end{eqnarray}
where $\sigma$ is the conductivity of the medium and $c$ is the velocity of 
light. For typical frequency and wave vector, this expansion parameter is
quite small, e.g. $10^{-4}$ for Fe ($\sigma=9\times10^{16}$ s$^{-1}$). 
Therefore, the series converges rapidly and only the leading term has 
practical interest. The $n-$th order terms in the expansion of 
Eq. (\ref{eq:series}) obey the 
Maxwell equations 
\begin{eqnarray} 
\nabla \cdot \mathbf{E}^{(n)} & = & 0 \nonumber \\
\nabla \cdot (\mathbf{H}^{(n)} + 4 \pi \mathbf{M} \delta_{n0} ) & = & 0 
\nonumber \\
\nabla \times \mathbf{E}^{(n)} & = & -\frac{1}{c} \frac{\partial \mathbf{B}^{(n)}}{\partial t} \nonumber \\
\nabla \times \mathbf{H}^{(n+1)} & = & \frac{4\pi\sigma}{c}  \mathbf{E}^{(n)} \nonumber \\ \label{eq:maxwell}
\end{eqnarray}
We neglect the displacement current in the last expression owing to $\omega \ll
\sigma$.
From Eq. (\ref{eq:maxwell})
$\nabla \times \mathbf{H}^{(0)}=0$, so we can write the  magnetic field as the
gradient of a magnetic scalar potential: 
$ \mathbf{H}^{(0)}=-\nabla \phi_{\mathrm{M}}$. 
The scalar potential has a volume and a surface term and the zeroth order 
magnetic field produced by the spin waves can be written as \cite{Jackson}
\begin{eqnarray}
\mathbf{H}^{(0)} & = & -\int_{V}d^{3}r'\frac{\mathbf{\nabla}\cdot\mathbf{M}(\mathbf{r'})(\mathbf{r}-\mathbf{r'})}{\left|\mathbf{r}-\mathbf{r'}\right|^{3}}  \nonumber \\
& + & \int_{S}d^{2}r'\frac{\hat{n}\cdot\mathbf{M}(\mathbf{r'})(\mathbf{r}-\mathbf{r'})}{\left|\mathbf{r}-\mathbf{r'}\right|^{3}}\label{eq:field}
\end{eqnarray}
where $\mathbf{M}$ is the magnetization of the sample and $\hat{n}$ is the 
outward normal to the surface carrying magnetic charge. 
We shall consider films to be thin enough that the spin waves are confined
 to the $x-y$ plane only. We shall consider two specific cases: 
$\mathbf{k} \bot \mathbf{M}$ and $\mathbf{k} \| \mathbf{M}$.

We consider the spin wave propagating along the $x$ direction in the thin film.
In configuration I, the magnetization is precessing in the $x-z$ plane
\[
\mathbf{M}^{\mathrm{I}}=M_{0}\widehat{y}+\epsilon\left[\cos\left(kx-\omega t\right)\widehat{x}+\sin\left(kx-\omega t\right)\widehat{z}\right]
\]
where $M_{0}$ is the component of magnetization perpendicular to the plane of 
precession and $\epsilon$ is the amplitude
of precession. In our calculation, we need not restrict $\epsilon$ to be 
small compared to $M_0$.
The zeroth order magnetic field from Eq. (\ref{eq:field}) for this configuration 
is
\begin{align*}
H_{x}^{\mathrm{I}(0)}&  = 4\epsilon\pi\cos(kx-\omega t)\left[e^{-kz}-1\right]
 \; & \mathrm{for}\:  k & >  0\\
  &= 4\epsilon\pi\cos(kx-\omega t)\left[e^{-k(z-d)}-1\right]  
\; &\mathrm{for}\:   k & <  0\\
H_{y}^{\mathrm{I}(0)}  &= 0 \quad & \quad &\\
H_{z}^{\mathrm{I}(0)}  &= -4\epsilon\pi\sin(kx-\omega t)e^{-kz}\,\;
& \mathrm{for}\: k & >0\\
  &= -4\epsilon\pi\sin(kx-\omega t)e^{-k(z-d)}\,\; & \mathrm{for}\: k &< 0
\end{align*}

According to Eq. (\ref{eq:maxwell}) $\mathbf{H}^{\mathrm{I}(0)}$ generates an
electric field $\mathbf{E}^{\mathrm{I}(0)}$ 
which has only one nonzero component

\begin{align}
E_{y}^{\mathrm{I}(0)}&  =  \frac{2\xi}{k}\sin(kx-\omega t)[1-e^{-kz}]
\quad &\mathrm{for} \, k>0 \nonumber \\
E_{y}^{\mathrm{I}(0)}&  =  \frac{2\xi}{k}\sin(kx-\omega t)[1-e^{-k(z-d)}]
\quad &\mathrm{for} \, k<0 \nonumber \\
\label{eq:EI}
\end{align}
where $\xi = 2\pi\epsilon\omega/c$.
Note the asymmetry of the solution for positive and negative values of $k$. 
The profile of the electric field for $k>0$ and $k<0$ are mirror symmetric
with respect to $z=d/2$. 

In configuration II, the magnetization is precessing in the $y-z$ plane
\[
\mathbf{M}^{\mathrm{II}}=M_{0}\widehat{x}+\epsilon\left[\cos\left(kx-\omega t\right)\widehat{y}+\sin\left(kx-\omega t\right)\widehat{z}\right]
\]
Since $\nabla\cdot\mathbf{M}^{\mathrm{II}}=0$ only the surface term
of Eq. (\ref{eq:field}) contributes to the magnetic field given by
(for $k>0$)
\begin{eqnarray*}
H_{x}^{\mathrm{II}(0)} & = & -2\epsilon\pi\cos(kx-\omega t)[e^{k(z-d)}-e^{-kz}]\\
H_{y}^{\mathrm{II}(0)} & = & 0\\
H_{z}^{\mathrm{II}(0)} & = & -2\epsilon\pi\sin(kx-\omega t)[e^{k(z-d)}+e^{-kz}]
\end{eqnarray*}
which generates an electric field
\begin{eqnarray}
E_{x}^{\mathrm{II}(0)} & = & \left[-Ae^{kz}+Be^{-kz}\right]\sin(kx-\omega t)\nonumber \\
E_{y}^{\mathrm{II}(0)} & = & \frac{\xi}{k}\left[2-e^{k(z-d)}-e^{-kz}\right]\sin(kx-\omega t)\nonumber \\
E_{z}^{\mathrm{II}(0)} & = & \left[Ae^{kz}+Be^{-kz}-\frac{2\xi}{k}\right]\cos(kx-\omega t)\label{eq:EII}
\end{eqnarray}
where $A=(\xi/k)[(1-e^{-kd})/\sinh(kd)]$ and $B=(\xi/k)[(e^{kd}-1)/\sinh(kd)]$. Note that unlike in the previous configuration the components of
the electric field are symmetric with respect to the two surfaces.

\begin{figure}
\includegraphics[width=2.9in]{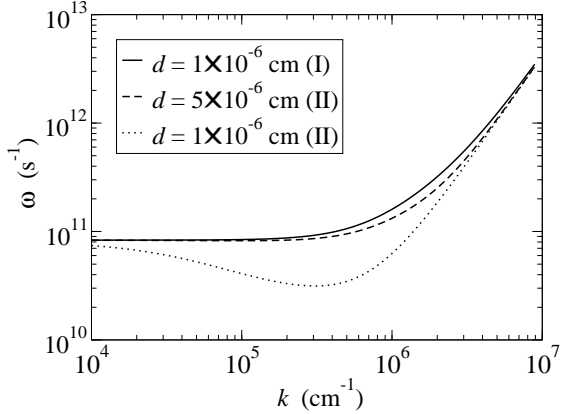}
\caption{\label{disp}Dispersion relation for configurations I and II
according to Eqs. (\ref{eq:omega}) 
and (\ref{eq:omegamag})
for an infinite iron slab using $M_S = 1700$
 emu/cc, $A$ = $2\times 10^{-6}$ erg/cm, $\gamma$ = $1.76\times 10^7$ 
(Oe s)$^{-1}$ and $H_{\mathrm{ext}}=1000$ Oe.}
\end{figure}

The energy stored in the form of spin waves is dissipated from the system 
by the current generated by
the electric field induced by the precessing spins. This induced electric 
field drives the free electrons in the metal to produce the magnon induced
current. The ohmic power loss per unit volume due to this magnon induced 
current can be written as
\[
P=\lim_{L\rightarrow\infty}\frac{\sigma}{2Ld}\int_{0}^{d}dz\int_{-L}^{+L}dx\left(E_{x}^{2}+E_{y}^{2}+E_{z}^{2}\right)
\]
Integrating over the square of the electric field
described in Eqs. (\ref{eq:EI}) and (\ref{eq:EII}),
we obtain the power dissipation per unit volume
\begin{eqnarray}
P^{\mathrm{I}}(k,d) & = & 2 \beta\left[-3+2dk+4e^{-dk}-e^{-2dk}\right]\nonumber \\
P^{\mathrm{II}}(k,d) & = & \frac{\beta e^{-2kd}}{(1+e^{kd})}[(-15+8dk)e^{3dk} 
\nonumber \\
& + &(9+10dk)e^{2dk}+(7+2dk)e^{dk}-1] \nonumber \\
\label{eq:power}
\end{eqnarray}
where 
\begin{eqnarray}
\beta & = & \frac{2\sigma\pi^{2}\epsilon^{2}\omega^{2}}{dk^{3}c^{2}} 
\label{eq:beta}
\end{eqnarray}

\begin{figure}
\includegraphics[width=2.9in]{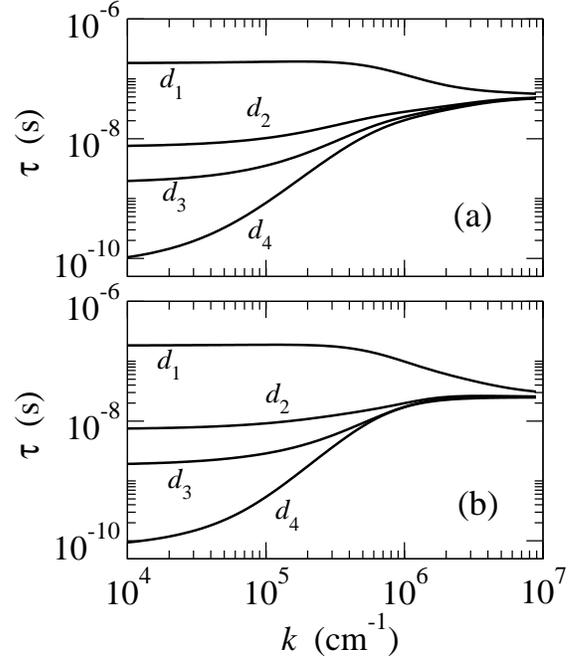}
\caption{\label{tauvsk}Relaxation time for an infinite iron thin film 
with thickness $d_1=1\times 10^{-6}$, $d_2=5\times 10^{-6}$, $d_3=1\times 10^{-5}$
and $d_4=5\times 10^{-5}$ cm
for (a) configuration I and (b) configuration II.}
\end{figure}

The power dissipation clearly depends on $\omega$ which is a function of $k$.
Owing to the strong influence of the magnetostatic energy within the magnon
Hamiltonian, the derivation of this relationship is nontrivial, but, fortunately
has been described by previous workers. Essentially, the
Hamiltonian of the system has contribution from exchange, magnetostatic
and Zeeman energy. We shall restrict ourselves to isotropic systems and 
therefore crystallographic anisotropy will have negligible effect to our
result. We will only consider magnons with wavelength much greater than the
lattice constant, e.g. $k \le 10^7$ cm$^{-1}$ which applies to most magnons of 
interest.
The dispersion relation for a thin film is given by \cite{S1970PRB}
\begin{equation}
\omega = \gamma \sqrt{\left(H_i+\frac{2A}{M_S}k^2 \right)\left(H_i+\frac{2A}{M_S}k^2 +2\omega_d\right)}
\label{eq:omega}
\end{equation}
where $H_i=H_{{\mathrm{ext}}}-NM_S$ is the internal field, $A$ is the exchange 
constant, $N$ is the demagnetizing factor and $H_{{\mathrm{ext}}}$ is the external 
magnetic field. The magnetostatic contribution is given by
\begin{eqnarray}
\omega_d^{\mathrm{I}} & = & 2\pi M_S \nonumber \\
\omega_d^{\mathrm{II}} & = & 2\pi M_S \left[ \frac{1-e^{-kd}}{kd}\right] 
\nonumber \\
\label{eq:omegamag}
\end{eqnarray}
The dispersion relation for
an iron thin film of various thicknesses is shown in Fig. \ref{disp}.
Note that the curves converge for $k\gtrapprox 2\times 10^6$ cm$^{-1}$ where
exchange interaction starts dominating the thickness dependent magnetostatic 
interaction. 

The energy density of the system consists of the exchange, the magnetostatic
and the Zeeman terms
\begin{equation}
\mathcal{E}(k,d)=A\left( \frac{\nabla M}{M_S} \right)^2+ 
\mathcal{E}_d
- \mathbf{M}\cdot\mathbf{H}_{\mathrm{ext}}
\label{eq:energy}
\end{equation}
where the magnetostatic contribution to the energy is obtained by
taking the negative scalar product of magnetization and the magnetic
field produced by the magnons and is given by
\begin{eqnarray}
\mathcal{E}_d^{\mathrm{I}} & = & 2\pi\epsilon^2  \nonumber \\
\mathcal{E}_d^{\mathrm{II}} & = &
2\pi\epsilon^2 \left[ \frac{1-e^{-kd}}{kd} \right]   \nonumber \\
\label{eq:energymag}
\end{eqnarray}
 Eqs. (\ref{eq:power}), 
(\ref{eq:beta}),  
(\ref{eq:omega}), (\ref{eq:omegamag}), 
(\ref{eq:energy}) and (\ref{eq:energymag})
show that the energy decays exponentially with relaxation time:
\[
\tau(k,d) = \frac{\mathcal{E}(k,d)}{P(k,d)}
\]
This is the leading order term contributing to the relaxation time as 
the energy of the system gets renormalized by the 
magnetic field generated by the magnon-induced currents. 
Fig. \ref{tauvsk} shows the relaxation time of spin waves in an iron thin film
where the spin wave is confined in the $x-y$ plane. 
According to \cite{DV2003PRL} the four magnon scattering 
produces magnons with wave vectors in the range 
$1-5\times 10^6$  cm$^{-1}$.  
It is important to note that magnetostatic is the dominant interaction in 
the long wavelength limit whereas exchange takes over in the short 
wavelength regime. The crossover, which happens around $k\sim 10^6$ cm$^{-1}$
in the range of thickness of the sample we are interested in, is of the same
order of magnitude for bulk magnetic crystals \cite{Mills}. This feature 
is nicely reproduced in our result where changing
the value of the exchange constant only affects the curves 
for $k\ge10^6$ cm$^{-1}$.  

Both for configurations I and II, $\tau$
increases with $k$ for thicker films in the magnetostatic
regime and saturates in the exchange regime. For such films
the electric field 
and hence the power shows an inverse dependence with the wave vector in the
long wavelength limit whereas the energy of the system is independent of $k$ in
configuration I and changes slowly
with $k$ in configuration II.  This makes $\tau$ an increasing function 
of $k$ except for small thicknesses of the film in which case the power 
dissipation becomes nearly independent of $k$. However, both $\mathcal{E}$ and $P$ 
increases as $k^2$ in the exchange regime thereby making $\tau$ independent
of $k$. 

\begin{figure}
\includegraphics[width=2.9in]{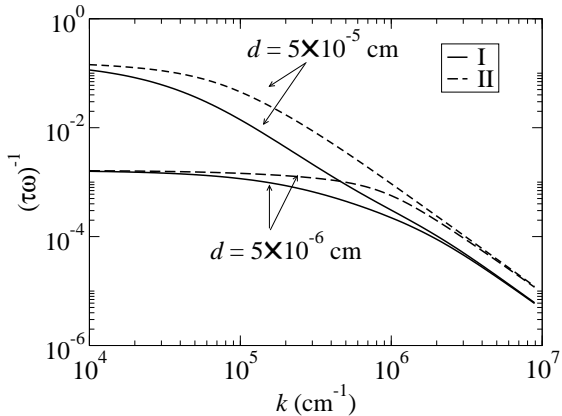}
\caption{\label{tauomega}The quantity $1/\tau\omega$ as a function of 
the wave vector for two different thicknesses.}
\end{figure}


Fig. \ref{tauomega} shows the behavior of the quantity $1/\tau\omega$
as a function of the wave vector. This quantity, which can be interpreted
as a wave vector dependent damping constant, is a measure of the amount of 
energy taken away from the system per precessional cycle. This damping constant
should not be confused with the typical Gilbert damping
constant used for uniform rotation of the element magnetization. 
However, the presence of values 
approaching $0.2$ illustrates the large magnitude of the effects described here.

Observation of spin wave resonance \cite{ST1958PRL}
in a thin film allows calculation of relaxation times from measured line widths. 
For example, analysis of Okochi's  \cite{O1970JPSJ} data for his first spin wave
resonance mode yields a damping constant $\alpha=0.0063$. Calculations for this 
configuration yield $P=32\pi^2\epsilon^2\omega^2\sigma/(c^2k^2) $ and 
$\mathcal{E}=(2Ak^2/M^2_S-4\pi+H_{\mathrm{ext}}/M_S)\epsilon^2$. The resulting value 
of the damping constant is $0.0095$ 
(using $A=10^{-6}$ erg/cm and 
$\sigma=2.9\times 10^{16}$ s$^{-1}$ \cite{PMW1966JAP} for FeNi).
The discrepancy is presumably due to conductivity differences between the two thin film
samples.
It is worth noting these experiments rely on exciting standing waves along the 
perpendicular direction of the thin film. This geometry 
typically minimizes the effect (relative to the configurations discussed elsewhere in this
text)
of magnon induced currents in the relaxation time because the magnon generated
magnetic field is zero which makes the induced electric fields weaker and the
rate of energy dissipation slower.

The effect of conductivity on the magnon-electron dissipation mechanism
can be studied in magnetic semiconductors such as CdCr$_2$Se$_4$ and 
HgCr$_2$Se$_4$ whose conductivity can be tuned by the amount of Ag doping
\cite{FC1978SSC,FC1996PRB}. The typical conductivity of such materials 
is several orders of magnitude lower than that of a ferromagnetic metal.
For example, a 0.75 mole $\%$ Ag doped CdCr$_2$Se$_4$ has $\sigma=4.5 \times
10^{11}$ s$^{-1}$ at $T=120 $K. Therefore, the coefficient of expansion in Eq. 
(\ref{eq:series}) becomes much smaller compared to that for a ferromagnetic metal
thereby making our formalism highly
applicable for such materials.  Assuming a resonance field 
$H_{\mathrm{res}}= 3500$ Oe \cite{FC1996PRB} and $\omega=\gamma  H_{\mathrm{res}}$ 
we obtain a linewidth $\Delta H=1/(\tau \gamma) = 78$ Oe for 
configuration I and $\Delta H = 144$ Oe for configuration II, using $d=0.2$ mm 
 and $k=1/d$. These values are in very good
agreement with the FMR linewidths observed by Ferreira and Coutinho-Filho
\cite{FC1978SSC}. Under the experimental condition the exchange contribution 
to the energy is negligible and the magnetostatic and the Zeeman energies are 
of the same order of magnitude.	In this limit, therefore, the magnon-electron 
contribution to the energy dissipation (calculated here) is comparable to that 
of conventional
Eddy current loss obtained from the FMR linewidth by subtracting the
effect of two magnon scattering \cite{Sparks1967IBID}. 

We conclude by proposing the following picture of ferromagnetic relaxation 
in switching experiments. We expect that the initial rapid approach of
magnetization direction to equilibrium is enabled by magnon-magnon scattering 
that converts the energy into the higher spin wave modes. These modes then 
decay at a slower pace via the magnon-electron interaction described here or
by the traditionally invoked mechanisms in less pure, lower conductivity films. 
This delay will lead to a small reduction in magnetization which appears to
have been observed by Silva et al \cite{SKP2002APL}.

This work was supported primarily by the MRSEC Program of the 
National Science Foundation under Award Number DMR-0212302. We thank Paul
Crowell and Alexander Dobin for useful discussions.


\begin{thebibliography}{1}

\bibitem{HSF1997PRL}W. K. Hiebert, A. Stankiewicz and M. R. Freeman, Phys.
Rev. Lett. \textbf{79}, 1134 (1997).
\bibitem{Choi2001PRL}B. C. Choi, M. Belov, W. K. Hiebert, G. E. Ballentine, and M. R. Freeman, Phys. Rev. Lett. \textbf{86}, 728 (2001).
\bibitem{AB2000Sci}Y. Acremann and C. H. Back, Science \textbf{290}, 492 (2000).
\bibitem{DV2003PRL}A. Y. Dobin and R. H. Victora, Phys. Rev. Lett. \textbf{90}, 
167203 (2003).
\bibitem{A1955PR}E. Abrahams, Phys. Rev. \textbf{98}, 387 (1955).
\bibitem{KP1975PRB}V. Kambersky and C. E. Patton, Phys. Rev. B. \textbf{11}, 
2668 (1975) and references therein.
\bibitem{AM1996PRB}N. S. Almeida and D. L. Mills, Phys. Rev. B. \textbf{53}, 
12232 (1996).
\bibitem{HP1940PR}T. Holstein and H. Primakoff, Phys. Rev. \textbf{58}, 
1098 (1940).
\bibitem{Jackson}J. D. Jackson, \emph{Classical Electrodynamics} (John Wiley
and Sons, New York, 1975).
\bibitem{HP1940PRB}T. Holstein and H. Primakoff, Phys. Rev. \textbf{58}, 
1098 (1940).
\bibitem{S1970PRB}M. Sparks, Phys. Rev. B. \textbf{1}, 3831 (1970).
\bibitem{Mills}D. L. Mills, \emph{Surface Excitations}, Modern Problems in 
Condensed Matter Science Vol. 9, edited by V. M. Agranovich and R. Loudon 
(Elsevier, Amsterdam, 1984).
\bibitem{ST1958PRL}M. H. Seavey and P. E. Tannenwald, Phys. rev. Lett. 
\textbf{1}, 168 (1958).
\bibitem{O1970JPSJ}M. Okochi, J. Phys. Soc. Jap. \textbf{28}, 897 (1970).
\bibitem{PMW1966JAP}C. E. Patton, T. C. McGill and C. H. Wilts, J. App. Phys.
\textbf{37}, 3594 (1966).
\bibitem{FC1996PRB}J. M. Ferreira and M. D. Coutinho-Filho, Phys. Rev. B. 
\textbf{54}, 12979 (1996).
\bibitem{FC1978SSC}J. M. Ferreira and M. D. Coutinho-Filho, Solid State Comm.
\textbf{28}, 775 (1978).
\bibitem{Sparks1967IBID} M. Sparks, J. Appl. Phys. \textbf{38}, 1031 (1967).
\bibitem{SKP2002APL}T. J. Silva, P. Kabos and M. R. Pufall, Appl. Phys. Lett.
\textbf{81}, 2205 (2002).

\end{thebibliography}
\end{document}